\documentclass{eas}

\usepackage{graphicx}

\begin{document}

%%-----------------------------

%%      the top matter

%%-----------------------------

\title{About the evolution of Dwarf Spheroidal Galaxies}
\runningtitle{About the evolution of dSphs}
\author{Andrea Marcolini}\address{Institute of Astronomy, Vienna University,
        T\"urkenschanzstrasse 17, 1180 Vienna, Austria}
\author{Annibale D'Ercole}\address{INAF - Osservatorio Astronomico di Bologna,
        via Ranzani 1, 40127 Bologna, Italy}
\author{Fabrizio Brighenti}\address{Dipartimento di Astronomia, Universit\`a
        di Bologna, via Ranzani 1, 40127 Bologna, Italy}
\author{Simone Recchi$^{ 1,}$}\address{INAF - Osservatorio Astronomico di 
        Trieste, via Tiepolo 11, 34131, Trieste, Italy}
%\author{ }\address{}
%

%

\begin{abstract}
  We present 3D hydrodynamic simulations aimed at studying the
  dynamical and chemical evolution of the interstellar medium in dwarf
  spheroidal galaxies. This evolution is driven by the explosions of
  Type II and Type Ia supernovae, whose different contribution is
  explicitly taken into account in our models. We compare our results
  with avaiable properties of the Draco galaxy. Despite the huge
  amount of energy released by SNe explosions, in our model the galaxy
  is able to retain most of the gas allowing a long period ($> 3$ Gyr)
  of star formation, consistent with the star formation history
  derived by observations. The stellar [Fe/H] distribution found in
  our model matches very well the observed one. The chemical properties
  of the stars derive from the different temporal evolution between
  Type Ia and Type II supernova rate, and from the different mixing of
  the metals produced by the two types of supernovae. We reproduce
  successfully the observed [O/Fe]-[Fe/H] diagram.
\end{abstract}

\maketitle

%%-----------------------------
%%      your text
%%-----------------------------
\section{Introduction}

Due to their proximity, galaxies of the Local Group (see Mateo 1998,
Grebel 2006 for a review) offer an unique opportunity to study in
detail their structural properties, formation and chemical evolution.

In particular, the distribution of the local galaxies shows the
clustering of dwarf ellipticals and dwarf spheroidal (dSphs) around
the dominant spirals galaxies (Milky Way and Andromeda).  Dwarf
spheroidals are the least massive galaxies known, but yet, their
velocity dispersions imply mass to light ratios as large as 100 $\rm
M_{\odot}/L_{\odot}$. This is usually explained assuming that these
systems are dark matter dominated. Actually, in the past few years
both observational evidences (e.g. Kleyna et al. 2002, Lokas 2002, 
Walker et al. 2006) and
theoretical works (e.g. Kazantzidis et al.  2004, Mashchenko et al.
2005) confirm the possibility that these galaxies are relatively
massive bounded system with virial masses in the range
$10^8-5\times10^9$ M$_{\odot}$.  Such galaxies are very metal poor and
lack of neutral hydrogen and recent star formation. Thus they were
initially believed to be very similar to Galactic globular clusters
and to have a very simple star formation history (SFH).  Recent
studies have shown, instead, that these systems are much more complex,
with varied and extended SFHs.  High resolution spectroscopy of
several dSphs showed the presence of a wide range in metallicity
(Harbeck et al. 2001).  For example, abundance analyses of stars
belonging to Draco and Ursa Minor have shown values of [Fe/H] in the
range $-3 \leq$[Fe/H]$\leq -1.5$ (Shetrone et al. 1998, 2001) with a
mean value in the interval $-2.0 \leq \langle$[Fe/H]$\rangle \leq
-1.6$, depending on the authors (e.g.  Shetrone et al. 2001,
Bellazzini et al. 2002).

The above ranges are consistent, for some dSphs, with a single period
of star formation extended in time for a few Gyr (e.g. Mateo 1998,
Dolphin 2002). As a further hint of long SFH Shetrone et al. (2001)
found that their observed dSphs have [$\alpha$/Fe] abundances that are
$\sim 0.2$ dex lower than those of Galactic halo field stars in the
same [Fe/H] range. This suggests that the stars in these systems were
formed in gas pre-enriched by Type II supernovae (SNe II) as well as
by Type Ia supernovae (SNe Ia), and star formation must thus continue
over a relatively long timescale in order to allow a sufficient
production of iron by SNe Ia.

Given the small dynamical mass inferred for dSphs, the interstellar
medium (ISM) binding energy is small when compared to the energy
released by the SNe II explosions occurring during the star formation
period; for instance, as shown in the next section, in a dSph like
Draco the baryonic matter has a binding energy of $\sim 10^{53}$ erg,
while the expected number of SN explosions in the past was
$10^3-10^4$, realising an energy much larger than the binding energy.
It is thus quite puzzling how the ISM can remain bound long enough to
allow such a long star formation duration. Infact, contrary to SNe II,
SNe Ia are poor producers of oxigen, great producers of iron, and
start to explode after longer time scales; thus stars with low [O/Fe]
indicate long SFHs.

Motivated by the above arguments, in this paper we explore the
possibility that dSphs formed stars at a low SFR for a long period.
To compare our results with observations, we have tailored our models
on the Draco galaxy: this galaxy is supposed to have experienced a
star formation lasting for 3-4 Gyr, and which essentially ceased 10
Gyr ago (e.g. Mateo 1998). Obviously, our results may be confronted
with other dSphs which are strongly dark matter dominated and have
similar SFHs as, e.g., Ursa Minor (Mateo 1998).

We run a number of three-dimensional (3D) hydrodynamical simulations
to study the dynamical and chemical evolution of this system,
following an assumed SFH. A special attention is
paid to the influence of both SNe Ia and SNe II on the chemical
enrichment of the new forming stars.
 
\section{Model and discussions}

We start our simulations with the ISM in
hydrostatic equilibrium in the dark matter halo potential well.
Although the stellar contribution to the gravitational potential well
is neglected, we approximate the observed stellar distribution in Draco
with a King profile with a mass content $M_{*}=5.6 \times 10^5$
M$_{\odot}$. The stellar distribution is important to estimate (from
the observed mass-to-light ratio) the dark matter halo properties and
properly locate the SNe explosions. More details on the model
construction can be found in Marcolini et al 2006. One of the basic
assumptions of the model is that the dark matter halo extends beyond
the steller component ($R_{*}=650$ pc) of the system (the mass of
the dark matter halo at 1.2 kpc is $6.2 \times 10^7$ M$_{\odot}$). The
initial gas mass is $M_{\rm ISM} = 0.18 M_{\rm h}$, which corresponds
to the baryonic fraction given by Spergel et al. (2006).

\begin{figure}
  \centering
  \includegraphics[width=13cm]{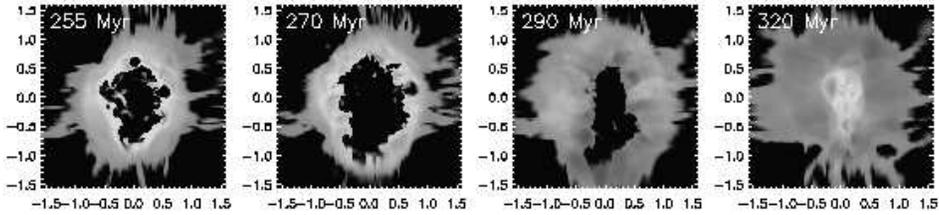}
  \caption{Logarithm of the density distribution (g cm$^{-3}$) of the
    ISM in the $z=0$ plane at different times. The first, second,
    third and fourth panels represent snapshots of the gas after a
    time interval $\Delta t=15$ Myr, 30 Myr, 50 Myr and 80 Myr from
    the occurrence of the latest instantaneous burst. Distances are
    given in kpc. The grey scale map ranges from -26.5 (black) to -24.5
    (white).}
\end{figure}

Here we focus on a model in which we assumed that stars form in a
sequence of 25 instantaneous bursts separated in time by 120 Myr. We
further assume that a single SN II explodes for each 100 M$_{\odot}$ of
formed stars, raching the total number of 5600 at the end of the simulation
(3 Gyr). The SNe II explode at a constant rate for 30 Myr (the
lifetime of a 8 M$_{\odot}$ star, the less massive SN II progenitor)
after the occurrence of each burst, while SNe Ia rate follows the
prescription of Matteucci \& Recchi 2001 (see Marcolini et al. 2006
for further details).

\begin{figure}
  \centering
  \includegraphics[width=11cm]{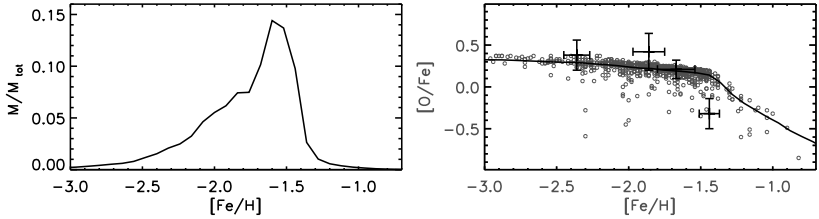}
  \caption{Left panel: final [Fe/H] distribution function of the long-lived
    stars. Right panel: abundance ratio [O/Fe] plotted against [Fe/H]
    of 1000 sampled stars. The solid line represents the mean value of
    the [O/Fe] distribution for any fixed [Fe/H]. We also show the
    observative values obtained by Shetrone et al (2001) for Draco.}
\end{figure}

Figure 1 shows that as the SNe II start to explode, a large fraction
of the central volume is filled by the hot rarefied gas of the SNRs'
interior, while the dense SNRs' shells form dense cold filaments after
colliding one with another. Once the SNe II stop to explode the global
cavity collapses and the ISM goes back into the potential well; this
happens nearly 30-40 Myr after the last SN II explosion. Note that the
initial binding energy of the gas $E_{\rm bind} \sim 8.3 \times
10^{52}$ erg is lower than the total energy $2.24 \times 10^{53}$ erg
released by the SNe II after a single burst. The simulation thus shows
that the radiative losses are substantial and prevent the evacuation
of the gas, as shown in Fig. 1.

After 120 Myr a central high density gas region of the same size of
the stellar volume is recovered, although turbulences and
inhomogeneities are now present (see Fig. 1, fourth panel). A second
burst of star formation then occurs leading to a second sequence of SN
II explosions. The gas undergoes a new cycle of merging bubbles which
eventually collapse again. 

The influence of the SN Ia explosions on the
general hydrodynamical behaviour of the ISM is not very important
because during a cycle of SN II explosions no more than 8-9 SNe Ia
occur, only $\sim 4\%$ of the SN II number. Despite their little
importance from a dynamical point of view, the role of SNe Ia is very
relevant for the chemical evolution of the stars. The simulation shows
that while the SN II ejecta become more and more homogeneous with time
as the turbulence diffuse it, the SN Ia ejecta appear to be
distributed less homogeneously; the reason for this is the low SNe Ia
rate.

We point out that during the entire evolution the fraction of the
SN ejecta present inside the stellar region remains very low ($\sim 18\%$
after 3 Gyr).  This is the amount of metals which contributes to the
metallicity of the forming stars. A large fraction of the ejecta is
pushed at larger distances by the continuous action of the SN
explosions. Figure 2 shows the [Fe/H] distribution function (MDF) of
our simulated long-lived stars (with mass $\le$ 0.9 M$_{\odot}$), i.e.
the mass fraction of these stars as a function of their [Fe/H].  At
the end of the simulation we obtain a mean value of
$\langle$[Fe/H]$\rangle$=-1.7 with a spread of $\sim$ 1.5 dex, in
reasonable agreement with observations (e.g. Shetrone et al. 2001,
Bellazzini et al. 2002), while the distribution maximum occurs at
[Fe/H]$\sim$-1.6. Note that stars with [Fe/H] $\ge$ -1.4 (the high
metallicity tail in Fig. 2) are particularly enriched by SN Ia ejecta
and formed in the (relatively) small volume occupied by SN Ia
renmants. This is particular evident in Fig. 2 (right panel) where we
show the final [O/Fe]-[Fe/H] diagram. The open circles form a
statistically representative sample of the stellar distributions in
the [O/Fe]-[Fe/H] diagram. The plateau at [O/Fe]$\sim$0.35 at low
[Fe/H] is representative of the [O/Fe] value in SNe II ejecta, because
the contribution of SNe Ia becomes important after a longer time
scale.  Indeed the small negative gradient of the plateau is due to
the slowly growing contribution in the Fe enrichment by SNe Ia (which
contribute only marginally to the Oxigen production). The sharply
decreasing branch at higher [Fe/H] is due to stars formed in the
regions of ISM recently polluted (mostly by iron) by SNe Ia.  A glance
at Fig. 2 shows that the stars on the decreasing branch populate the
MDF high [Fe/H] tail, while the majority of the stars occupies the
high [Fe/H] edge of the plateau in the [O/Fe]-[Fe/H] diagram. We point
out that our representative stellar sample is in reasonable agreement
with the stars observed by Shetrone et al. (2001).

\subsection{Other models}

Here we describe the evolution of two models quite similar
to the reference model (described above). Model B has the same SFH but
differs in the dark matter content ($2.2 \times 10^7$ M$_{\odot}$) and
ISM mass ($4 \times 10^6$ M$_{\odot}$) in order to preserve the cosmological
ratio between the amount of baryonic and non-baryonic matter.
Model C has the same properties of the reference
model but differs in the duration ($\le$ 1 Gyr) and intensity (10
bursts) of SFH.

Model B loses all its gas in a period too short ($\le$ 250 Myr) to be
consistent with the longer SFH of Draco. We point out that this
effective gas removal is mainly due to a less efficient radiative
cooling (due to the lower density of the gas) rather than to the
shallower galactic potential.

Model C, instead, retains its gas for a longer time, and is able to
form stars up to 900 Myr, consistently with recent cosmological
simulations (e.g. Ricotti \& Gnedin 2005, Kawata et al. 2006) before
loosing the ISM via a galactic wind. This model shows $\alpha$/Fe
$\sim$ 0.1 dex higher than the value of the reference model because of
the lower number of SN Ia explosions. However, although its chemical properties
(both the AMD and the [O/Fe]-[Fe/H] diagram) are still in marginal
agreement with observations, we belive that the star formation
duration must be longer than $\sim 1$ Gyr.
%Indeed, Shetrone et al. found that dSphs
%have [$\alpha$/Fe] ratios $\sim0.2$ dex lower than those of the
%Galactic halo fields stars in the same [Fe/H] range, whose
%[$\alpha$/Fe] ratios are 0.3 (Shetrone et al. 2001).
Infact, Fenner et al (2006) find that only long SFHs (of the order of
several Gyr) are able to reproduce the Ba/Y ratio because the stars
must form over an interval long enough for the low-mass stars to
pollute the ISM with $s$-elements.

\section{Conclusion}

We presented 3D simulations of a forming dSph resembling the Draco
galaxy.  With our assumptions, in our reference model the galaxy never
gets rid of its gas (due to the huge efficiency of radiative cooling
despite the low metallicity of the gas) and the star formation can
last for several Gyr (as suggested by observations). This in turn
implies the need of an external mechanism to remove the gas and stop
the star formation, such as gas stripping (e.g. Marcolini et al 2003,
only stripping, and 2004, stripping+SN feedback) and/or tidal
interaction with the Galaxy (Mayer et al. 2006).  Indeed, ram pressure
due to gaseous haloes of the Milky Way and M31 is belived to explain
the observed correlation between stellar content and galactocentric
distance of dwarf galaxies (van den Bergh 1993).  We
are now running simulations of forming dSphs interacting with the Milky Way
halo in order to understand whether the combined action of SNe
feedback and ram pressure stripping can help in depriving these
systems of gas.

Although the SN ejecta remain gravitationally bounded during the star
formation, yet only a low fraction ($\sim 18$\%) stays in the region
where star forms. This effect mimics the
assumption of metal removal by galactic winds in chemical evolution
models. Our model succeeds in reproducing the [Fe/H] distribution
of the stars. In agreement with observations, we find a mean
value $\langle$[Fe/H]$\rangle = -1.7$ with a spread of $\sim 1.5$ dex.
We can also satisfactory reproduce the observed [O/Fe] vs [Fe/H]
diagram. The origin of the break in this diagram, in our
interpretation, is due to the low value of the porosity of SN Ia
remnants.  Indeed, given the low SN Ia rate, these remnants are located
quite apart one from another, and the iron ejected by SNe Ia is
distributed rather inhomogeneously through the stellar volume. As a
consequence, stars forming in the (relatively small) volume occupied
by SN Ia remnants have a ratio [O/Fe] lower than those forming
elsewhere.

%%-----------------------------
%%      your bibliography
%%-----------------------------

\end{document}